\documentclass[12pt]{article}
\usepackage{graphicx}
\usepackage{url}
\usepackage{pstricks}
\usepackage{color}
\usepackage{epsfig}
\usepackage{fancyhdr}
\usepackage{mathbbol}
\usepackage{pstricks}
\usepackage{color}
\usepackage{bbm}
\usepackage{multirow}
\usepackage{bm}
\usepackage{amsmath}
\usepackage{amsfonts}
\usepackage{amssymb}
\def\be{\begin{equation}}
\def\ee{\end{equation}}
\def\bea{\begin{eqnarray}}
\def\eea{\end{eqnarray}}

\title{Neutron Production Rates by Inverse-Beta Decay  in Fully Ionized Plasmas}

\author{
L. Maiani, A.D. Polosa and V. Riquer\\
{\small Dipartimento di Fisica and INFN, Sapienza Universit\`a di Roma,} \\
{\small Piazzale Aldo Moro 2, I-00185 Roma, Italy}
}

\begin{document}
\maketitle
\begin{abstract}
Recently we showed that the nuclear transmutation rates are largely overestimated in the Widom-Larsen theory of the so called `Low Energy Nuclear Reactions'. Here
we show that unbound plasma electrons are even less likely to initiate nuclear transmutations. 
\end{abstract}

\pagenumbering{arabic}

\subsection*{Introduction}

Claims of electron-proton conversion into a neutron and a neutrino by inverse beta decay in metallic hydrides have recently been raised~\cite{wi,wi2}, in the context of the so called Low Energy Nuclear Reactions (LENR). The condition for the reaction to occur is a considerable mass renormalization of the electrons, to overcome the negative Q-value that, otherwise, would forbid the reaction to occur. Defining a dimensionless parameter, $\beta$, in terms of the electron effective mass, $m^\star$~\footnote{To avoid confusion, we underscore that  the mass renormalization in~(\ref{massren}) has nothing to do with the velocity dependent relativistic mass. We consider extremely  non-relativistic  electrons. The situation is closely analogous to  muon capture in muonic atoms, in that case $m^\star$ being replaced by  the muon mass.}, one needs:
\be
\beta= \frac{m^\star}{m}\geq \frac{m_n -m_p}{m}\approx 2.8
\label{massren}
\ee
a reference value, $\beta=20$ was estimated in~\cite{wi}. 

It is not clear at all if such spectacularly large values of $\beta$ can be obtained in metallic hydrides and under which conditions. Nonetheless, {\it assuming} a given value of $\beta$, a calculation of the neutron rate can be obtained in a straightforward fashion from known electroweak physics. A calculation along these lines has been presented in Ref.~\cite{noi} for the case of an electron bound to a proton, superseding the order-of-magnitude estimate presented in~\cite{wi}.

More recently, the authors of Ref.~\cite{wi2} have argued that nuclear transmutations should most likely be started by unbound plasma electrons. Assuming a fully ionized plasma and completely unscreened electrons, they find a rate which is enhanced, with respect to the value obtained for bound electrons, by the so-called Sommerfeld factor, $S_0$ ($c=1$):
\be
S_0=\frac{2\pi \alpha}{v}
\label{sommer0}
\ee
where $\alpha$ is the fine structure constant and $v$ is the average thermal velocity of the electrons defined by\footnote{We shall use the numerical values: $k=8.617 ~10^{-5}$eV/$^0$K, $e^2 / \hbar c=\alpha=1/137.043$ and set $c=\hbar c=1$.}:
\be
v_{\rm th}=\sqrt{\frac{3kT}{m^\star}}=\beta^{-1/2}\sqrt{\frac{3kT}{m}}=3.6\cdot 10^{-4}\left[ \left(\frac{T}{5\cdot 10^3~^0{\rm K}}\right)\left(\frac{20}{\beta}\right)\right]^{1/2}
\label{vthermal}
\ee
with the numerical value in correspondence to $\beta=20$ and to the temperature $T \approx 5 \cdot 10^{3}~^0$K, estimated in~\cite{wi2} as the temperature that can be reached by hydride cathodes.
However, the assumption of completely unscreened electrons may be unrealistic. We consider here the situation in presence of  Debye screening, which, in a different context, has been recently analysed in Ref.~\cite{ArkaniHamed:2008qn}. We find  that at large densities, the plasma enhancement saturates to a value determined by the Debye length, $a_D$:
\be
S_0 \to S=\frac{a_D}{a^\star_B}
\label{sommerplas}
\ee
with:
\be
a^\star_B=\frac{1}{\alpha m^\star}=\beta^{-1}a_B
\ee
and $a_B$ the Bohr radius. 
\subsection*{Debye Length}
Static charges are screened in a plasma. The potential of the electric field
of a test charge at rest in a plasma is (in Gaussian units)
\be
\phi=\frac{e}{r}e^{-r/a_D}
\ee
where $a_D$ is the Debye length defined by:
\be
\frac{1}{a_D^2}=\frac{1}{a_e^2}+\frac{1}{a_i^2}
\ee
The two lengths $a_{e,i}$ are associated to electrons and ions respectively and are
given by~\cite{ll10}
\be
a_e =\left(\frac{kT_e}{4\pi n_e e^2}\right)^{1/2}
\label{debyeelec}
\ee
and:
\be
a_i =\left(\frac{kT_i}{4\pi n_i(Ze)^2}\right)^{1/2}
\label{debyeion}
\ee
The difference in temperature between electrons and ions is expected to occur
naturally because of the large difference of mass which impedes the exchange
of energy in electron-ion collisions. Here we will make the approximation
$a_D = a_e
$, which leads to the numerical value:

\be
a_D =4.87{\rm \AA}\times  \left [ \left (\frac{T}{5000^0{\rm K}}\right)  \left (\frac{ 10^{20}{\rm cm}^{-3}}{n_e} \right)\right]^{1/2}
\label{dlength}
\ee
or a Debye mass $m_D$:
\be
m_D=\frac{\hslash}{a_D}= 404~{\rm eV}
\ee

We therefore get a Debye length of about nine atoms (compared to $a_B = 0.5 $~\AA) in correspondence to the reference temperature $T \approx 5 \cdot 10^{3}~^0$K and a reference density $n_e=10^{20}$ cm$^{-3}$.  When considering the $n$ dependence, we shall restrict to the  range:
\be
 10^{14}~{\rm cm}^{-3}\leq n \leq 6\times 10^{23}~{\rm cm}^{-3}
 \label{rangeofden}
 \ee
Values between $10^8$ and $10^{14}$~cm$^{-3}$ are typical of glow discharges and arcs whereas a value of about $10^{22}$~cm$^{-3}$ is the free electron  density in Copper~\cite{am}.
 Around  $2.5\times 10^{21}~\text{cm}^{-3}$ the Debye length equals the Bohr radius\footnote{Electron capture occurs spontaneously during the formation of neutron stars, when the Fermi energy of the electrons increases above the threshold value, due to the gravitational pressure. This occurs at electron densities $\gtrsim 10^{31}$~cm$^{-3}$.}. 

\section*{Critical Velocity}

The Sommerfeld factors  in a plasma,  Eqs.~(\ref{screened}) and (\ref{unscreened}), can be obtained from an intuitive argument as follows (see the Appendix for a derivation from the Schr\"odinger equation following~\cite{ArkaniHamed:2008qn}).


We consider a critical value of the velocity, defined as:
\be
\frac{2 \pi \alpha}{v_{\rm crit}}=\alpha m^\star a_D
\label{criticalv}
\ee
In this condition, the de Broglie wavelength of the particle, is equal to the  Debye length\footnote{we use $\hbar=1$, so that $h=2\pi$.}:
\be
\lambda=\frac{2\pi}{m^\star v_{crit}}=a_D
\ee
For larger velocities, the 
wavelength is smaller and the particle probes a region of space smaller than $a_D$, where it sees an essentially unscreened Coulomb potential. In these conditions, we have to use  $S_0$, Eq.~(\ref{sommer0}). 

For smaller velocities, as $v \to 0$,  the 
wavelength gets larger than $a_D$. The Sommerfeld factor saturates to the value on the r.h.s. of (\ref{criticalv}) since the particle explores increasingly large portions of neutral plasma, and the screened Sommerfeld factor in Eq.~(\ref{sommerplas}) has to be considered.

The critical velocity defined by~(\ref{criticalv}) is:
\be
v_{\rm crit}=
2.48\cdot 10^{-4}\left(\frac{20}{\beta}\right)
\left(\frac{n}{10^{20}{\rm cm}^{-3}}\right)
\left(\frac{5000 ^0 {\rm K}}{T}\right)
\ee
We consider our electrons to be at $v_{\rm th}$, Eq.~(\ref{vthermal}). At the reference point, this is larger than $v_{\rm crit}$, hence we should apply the unscreened result, $S_0$. With increasing density, however, $v_{\rm crit}$ goes above $v_{\rm th}$ (at $n\sim 2\cdot 10^{20}$~{\rm cm}$^{-3}$) and one should apply the screened result, $S$.
  \section*{Transmutation Rates}
  
To translate the previous discussion into the expected rates for transmutation from electrons in a plasma, we first recall the rate  for the transmutation from bound electrons~\cite{noi}:
 \bea
&&\Gamma(\tilde{e}p\to n\nu_e)_{\rm bound}= |\psi(0)|^2\times \frac{1}{2 \pi}(G_F m_e)^2\left[1+3\left(\frac{g_A}{g_V}\right)^2 \right]
 \times (\beta-\beta_0)^2;\notag \\
 && |\psi(0)|^2=\frac{\beta^3}{\pi a_B^3} \notag \\
&& \Gamma_{\rm bound}[\beta=20]= 1.8\cdot 10^{-3}~{\rm Hz} 
\label{gammabound}
 \eea
The total rate is obtained by multiplying the result  $ \Gamma_{\rm bound}$  by the volume and  by the ion density, which we take equal to the electron density, $n$, because of global neutrality:
\be
{\rm Rate}_{\rm bound}= n\cdot V\cdot\Gamma_{\rm bound}
\label{ratebound}
\ee

In the case of plasma electrons, screened and unscreened rates are obtained by the substitution:
  \be
  |\psi(0)|^2\to n \cdot (S\; {\rm or}\;  S_0)
  \ee
and the rate is  proportional to $n^2$:
\be
{\rm Rate}_{\rm plasma}= n\cdot V\cdot\frac{\Gamma_{\rm bound}}{|\psi(0)|^2}\cdot n\cdot (S\; {\rm or} \; S_0)
\label{rateplasma}
\ee
$S$ and $S_0$ corresponding respectively to the screened Debye plasma and to the unscreened Coulomb case.

For convenience, we normalize the rates in plasma to the rate in Eq.~(\ref{ratebound}), computed for $\beta=20$, already a considerably large rate, although a factor of $\sim 300$ smaller than claimed in~\cite{wi}, and see if we can get anywhere close to unity or higher.

\begin{figure}[htb!]
\centering
\includegraphics[scale=1]{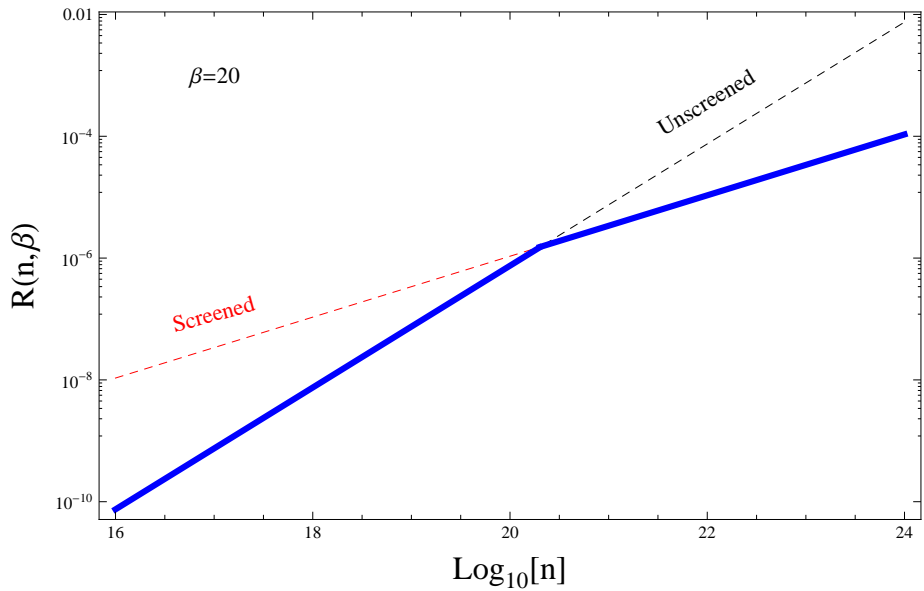}
\caption{\small  Ratios corresponding to the screened plasma (Sommerfeld factor $S$) and to the unscreened one (Sommerfeld factor $S_0$), for the case $\beta=20$. The previous discussion indicates that we must use $S_0$ for $v_{\rm crit}\leq v_{\rm th}$ and $S$ for $v_{\rm crit}\geq v_{\rm th}$. The result is represented by the thick line.}
\label{plasmaeff}
\end{figure}

The formulae are
\bea
\eta_{\rm Debye}(n,\beta)&=&\frac{{\rm Rate}_{\rm Debye}}{{\rm Rate}_{\rm bound}[\beta=20]}=n\frac{\pi a_B^3}{\beta^3}\frac{(\beta-\beta_0)^2}{(20-\beta_0)^2}S=\notag\\
&=&\frac{\pi (n a_B^3)}{\beta^2}\frac{a_D}{a_B}\frac{(\beta-\beta_0)^2}{(20-\beta_0)^2}
\label{ratdebye}
\eea
and
\be
\eta_{\rm Coul}(n,\beta)=\frac{{\rm Rate}_{\rm Coul}}{ {\rm Rate}_{\rm bound}[\beta=20]}=n\frac{2\pi\alpha}{v}\frac{\pi  a_B^3}{\beta^3}\frac{(\beta-\beta_0)^2}{(20-\beta_0)^2}
\label{ratcoulomb}
\ee
for the two cases.

In Fig.~\ref{plasmaeff} we display  the ratios corresponding to the screened plasma (Sommerfeld factor $S$) and to the unscreened one (Sommerfeld factor $S_0$), for the case $\beta=20$. The previous discussion indicates that we must use $S_0$ for $v_{\rm crit}\leq v_{\rm th}$ and $S$ for $v_{\rm crit}\geq v_{\rm th}$. The result is represented by the thick line.

The rate for electron capture from plasma never goes anywhere close to the capture rate for bound electrons derived in~\cite{noi} for the same value of $\beta$, let alone to the larger rate quoted in~\cite{wi}. Our results are in line with the lack of observation of neutrons in plasma discharge experiments recently reported in~\cite{enea}.  

\section*{Acknowledgements}
We thank Giancarlo Ruocco and Massimo Testa for interesting discussions.

\subsection*{APPENDIX: Sommerfeld factor for electrons in screened and unscreened  plasma}
Let us  consider an attractive screened potential in the plasma in the form:
\be
V (r) =- \frac{\alpha}{r} e^{m_D r} 
\label{screenpot}
\ee
The radial Schr\"odinger equation for the two body ($e^-$--ion) wave-function, $\chi(r)$, reads:
\be
\frac{d^2\chi(r)}{d^2 r}+ 2m^\star\left(m^\star \frac{v^2}{2}-V (r)\right)\chi(r)=0
\label{schreq}
\ee
Changing $r$  into  the adimensional variable $x$:
\be
r=a_B^\star x=\frac{1}{\alpha m^*}x
\label{scaled}
\ee
we get:
\be
\chi^{\prime \prime}(x)+\left(
\frac{v^2}{\alpha^2}+
\frac{2}{x} e^{-\epsilon x}
 \right)\chi(x)=0
\label{reduced}
\ee

In the limit of small or vanishing $v$  we write the equation as:
\be
\chi^{\prime \prime}(x)+k^2(x)\chi(x)=0
\label{reduced2}
\ee 
in terms of an effective momentum:
\bea
&& k^2(x)=\frac{2}{x} e^{-\epsilon x}
\eea
and solve it by the WKB method, which gives:
\be
\chi(x)=A
\frac{1}{\sqrt{k(x)}} e^{\pm i\int^x k(x')dx' }
\label{wkb}
\ee
We can use the WKB approximation as long as
\be
\left| \frac{k^\prime (x)}{k^2(x)} \right| \ll 1
\label{validity1}
\ee
that is:
\be
\frac{e^{\epsilon x/2}}{2\sqrt{2x}}(1+\epsilon x)\ll 1
\label{validity2}
\ee

At the value where the exponential bends, namely $\epsilon x =1$, we have:
\be
\left| \frac{k^\prime (x)}{k^2(x)} \right| _{x =1/\epsilon } =\sqrt{\frac{\epsilon}{2}}e^{1/2}=\frac{\epsilon}{k(x=1/\epsilon)}\equiv \frac{\epsilon}{k_{\rm eff}}
\ee
and the condition that this region is within the range of validity of WKB is then:
\be
\frac{v_{\rm eff}}{\alpha}=k_{\rm eff}\gg \epsilon= \frac{a_B^\star}{a_D}=\frac{ a_B}{\beta a}
\label{wkblimit}
\ee
with $\beta$ defined as in Eq.~(\ref{massren}). 

For $\beta =20$ and $a_D$ from Eq.~(\ref{dlength}), we find:
\be
v_{\rm eff}>\frac{\alpha a_B}{\beta a_D}\equiv v_{\rm WKB}\approx 3.9\cdot 10^{-5}
\ee
On the other hand, the smallest velocity we consider is the thermal velocity, Eq.~(\ref{vthermal}), which is safely within the region of validity of the WKB approximation. Note that $v_{\rm WKB}$ is simply proportional to the critical velocity $v_{\rm crit}$ defined in (\ref{criticalv}):
\be
v_{\rm WKB}=\frac{v_{\rm crit}}{2 \pi}
\ee

We are interested in the square modulus of the wavefunction at the origin relative to its unperturbed value (transmutation is taking place at the origin), the ratio being the  Sommerfeld enhancement: 
\be
S_k\sim |\psi_k(0)|^2 =\left|\frac{R_{k,\ell=0}(x=0)}{Ak}\right|^2 =\left|\frac{\chi_k(0)}{Axk}\right|^2
\label{essek}
\ee
where we have used the fact that $R_{k\ell}(x)\sim x^\ell$ as $x\to 0$. The constant $A$ depends on the normalization of the radial function at large distances~\footnote{In the conventions of~\cite{landau}, $A=2$}.  Since $R_{k,\ell=0}$ goes to a constant as $x\to 0$, we need that $\chi_k(x)\to 0$ as $x\to 0$ or
\be
\chi_k(x)\to x\chi_k^\prime(0)~\text{as}~x\to 0
\ee
thus giving 
\be
S_k\sim \left|\frac{\chi_k^\prime(0)}{Ak}\right|^2
\ee
Within the region of validity of the WKB approximation, $k\gtrsim \varepsilon$, we have  
\be
\chi(x)= A \frac{1}{\sqrt{k(x)}}e^{\pm i \int^x dx^\prime k(x^\prime)} 
\ee
where $A$ is chosen to be the same constant which appears in~(\ref{essek}). 
Therefore
\be
S_k\sim \left| \frac{1}{\sqrt{k(x)}}e^{\pm i \int^x dx^\prime k(x^\prime)} \left(\pm i-\frac{1}{2}\frac{k^\prime(x)}{k^2(x)}\right)\right|^2_{x=0}
\ee
the last term in parenthesis being much smaller than one. 
The {\it maximum} value attainable by  $S_k$ is at the border of the WKB approximation limit, {\it i.e.} for $k\sim \epsilon$, Eq.~(\ref{wkblimit})
\be
S\sim\frac{1}{\epsilon}= \frac{a}{a_B^\star}=\frac{a}{a_B}\beta
\label{screened}
\ee

In the limit $\epsilon \to 0$, the Schr\"odinger equation (\ref{schreq}) is solved analytically.
The $^\prime$in$^\prime$ wavefunction in the continuous spectrum of the attractive
Coulomb field is given by:
 \be
\psi^{(+)}_k= e^{\pi k/2}\Gamma(1- i/k)e^{i{\bf k\cdot r}}F(i/k,1,ikr-i{\bf k á r}) 
\label{coulombsol}
\ee
where $F=~_1F_1 $ 
is the Kummer function (hypergeometric confluent). Here ${\bf k\cdot r}$ 
corresponds to $mv\times r$, measured in units $1/m$. 
Thus it is the adimensional quantity $v/\alpha$. The same would hold writing $kr = (k/\alpha m)(\alpha m r)$.

In these respects $k/\alpha m\to k$ is dimensionless, $k = v/\alpha$, and we understand
the factor $e^{\pi k/2}$, or the term $\Gamma=(1- i/k)$. The $k = v/\alpha$ appears in the
Schr\"odinger equation (\ref{schreq}). 

The action of the attractive Coulomb field on the motion of the particle near the origin can be characterized by the ratio of the square modulus of
 $\psi^{(+)}(0)$ to the square modulus of the  wave function for free
motion $\psi_ k(r) = e^{i\bm k\cdot \bm r}$. Using that $\Gamma^*(z)$ =$\Gamma(z^*)$, $F(i/k, 1, 0) = 1$ and:

\be
\Gamma(1+i/k)\Gamma(1-i/k) =\frac{\pi}{k \sinh(\pi/k)}
\ee
we get the result:
\be
S=S_0=|\psi^{(+)}_{ k}(0)|^2=\frac{2}{k(1-e^{-2\pi/k})}
\approx \frac{2\pi}{k}=\frac{2\pi \alpha}{v}
\label{unscreened}
\ee
for small velocities~\cite{ArkaniHamed:2008qn,wi2}.

\end{document}